\documentclass{jfm}
\usepackage{graphicx}
\usepackage{epstopdf, epsfig}

\usepackage{amsmath,mathtools,amssymb,bm}
\usepackage{color}

\newcommand{\ue}{\mathrm{e}}
\newcommand{\ui}{\mathrm{i}}

\newcommand{\beq}{\begin{equation}}
\newcommand{\eeq}{\end{equation}}

\renewcommand{\p}{\partial}

\catcode`\@=11
\def\gtsim{\mathrel{\vcenter{\m@th\offinterlineskip
\hbox{$\hfill>\hfill$}\kern.5ex\hbox{$\hfill\sim\hfill$}}}}
\catcode`\@=12

\catcode`\@=11
\def\ltsim{\mathrel{\vcenter{\m@th\offinterlineskip
\hbox{$\hfill<\hfill$}\kern.5ex\hbox{$\hfill\sim\hfill$}}}}
\catcode`\@=12

\graphicspath{{./Figures/}}

\shorttitle{Flow in the cerebral aqueduct}
\shortauthor{S.~Sincomb and others}

\title{A model for the oscillatory flow in the cerebral aqueduct}

\author{%
S.~Sincomb\aff{1},
W.~Coenen\aff{1,2},
A.~L.~S\'anchez\aff{1}\corresp{\email{als@ucsd.edu}} \and
J.~C.~Lasheras\aff{1,3}}

\affiliation{%
\aff{1}Department of Mechanical and Aerospace Engineering, University of California San Diego, \\
La Jolla, USA
\aff{2}Grupo de Mec\'anica de Fluidos, Departamento de Ingenier\'ia T\'ermica y de Fluidos,
Universidad Carlos III de Madrid, Av.~Universidad 30, 28911 Legan\'es (Madrid), Spain
\aff{3}Department of Bioengineering, University of California San Diego, \\
La Jolla, USA}

\begin{document}

\maketitle

\begin{abstract}
This paper addresses the pulsating motion of cerebrospinal fluid in the aqueduct of Sylvius, a slender canal connecting the third and fourth ventricles of the brain. Specific attention is given to the relation between the instantaneous values of the flow rate and the interventricular pressure difference, needed in clinical applications to enable indirect evaluations of the latter from direct magnetic-resonance measurements of the former. An order-of-magnitude analysis accounting for the slenderness of the canal is used in simplifying the flow description. The boundary-layer approximation is found to be applicable in the slender canal, where the oscillating flow is characterized by stroke lengths comparable to the canal length and periods comparable to the transverse diffusion time. By way of contrast, the flow in the non-slender opening regions connecting the aqueduct with the two ventricles is found to be inviscid and quasi-steady in the first approximation. The resulting simplified description is validated by comparison with results of direct numerical simulations. The model is used to investigate the relation between the interventricular pressure and the stroke length, in parametric ranges of interest in clinical applications.

\end{abstract}


\section{Introduction}
\label{sec:introduction}

\begin{figure}
\centering
\includegraphics[width=1.0\linewidth]{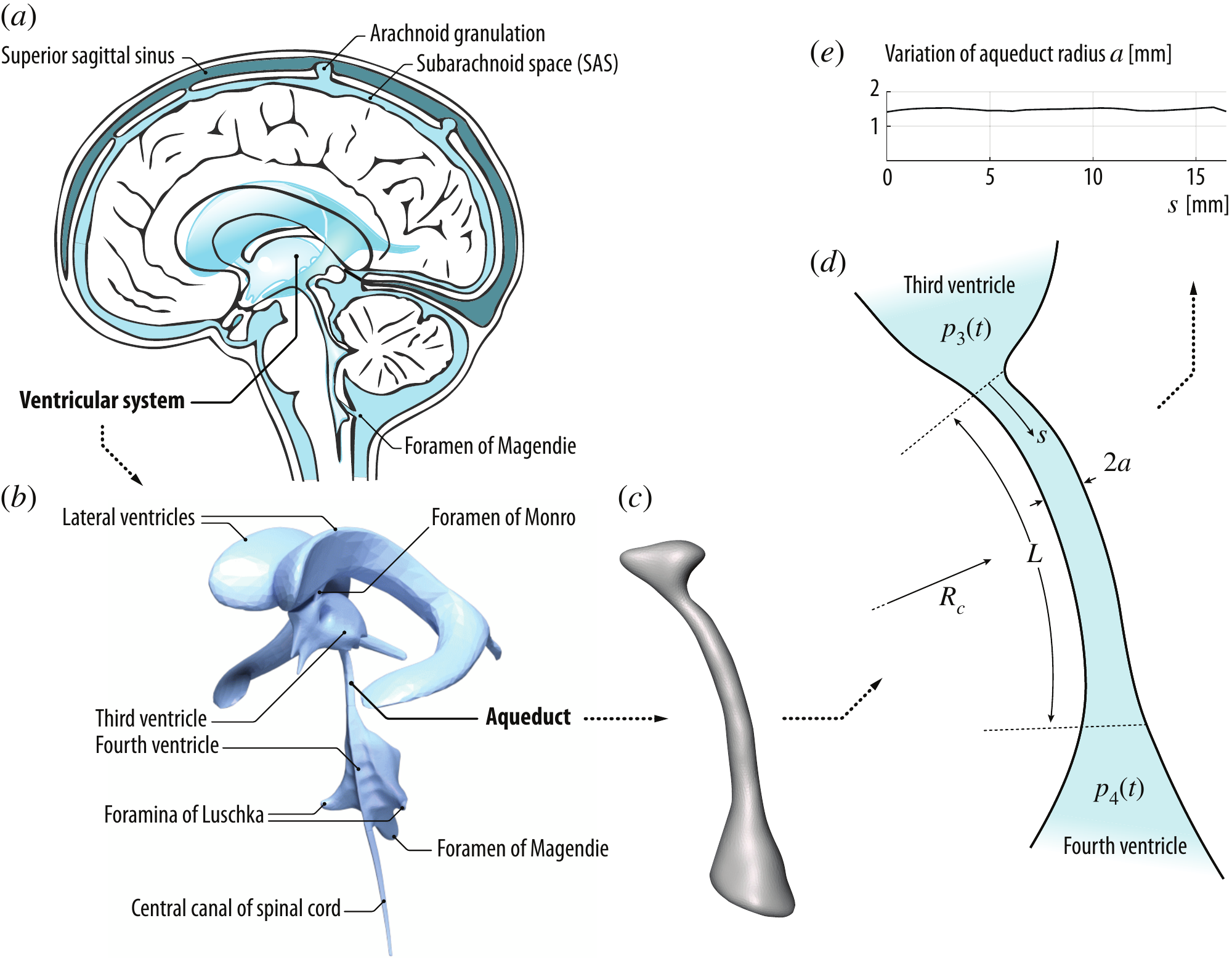}
\caption{(a) Schematic views of the cranial cavity  and (b) the cerebral ventricular system (BodyParts3D, © The Database Center for Life Science licensed under CC Attribution-Share Alike 2.1 Japan). Anatomic MR images are used to obtain (c) a smoothed surface mesh of the cerebral aqueduct of a healthy subject, which was used for (d) the simplified illustration highlighting the different flow regions and (e) the variation of the aqueduct radius with the distance to the third ventricle.}
\label{fig:1}
\end{figure}

The cerebrospinal fluid (CSF) is a colorless fluid with waterlike physical properties (i.e.\ density $\rho \simeq 10^3$ kg/m$^3$ and kinematic viscosity $\nu \simeq 0.71 \times 10^{-6}$ m$^2$/s) that bathes the central nervous system (CNS), filling the ventricles of the brain and the surrounding subarachnoid space (SAS), as shown in figure~\ref{fig:1}. The CSF flows between ventricles through their different interconnecting passages (or foramina), including the foramina of Monro, connecting the lateral ventricles with the third ventricle, the cerebral aqueduct (or aqueduct of Sylvius) connecting the third and fourth ventricles, and the foramen of Magendie and foramina of Luschka, connecting the fourth ventricle with the SAS. The resulting motion includes a steady component corresponding to the continuous flow from the ventricles, where CSF is secreted from the blood plasma in the choroid plexus, towards the SAS, where CSF is reabsorbed into the venous circulation at fingerlike projections of the arachnoid membrane surrounding the brain, called villi. Besides this slow steady motion, the CSF is known to undergo a much faster pulsating motion driven by the cardiac and respiratory cycles, with peak volumetric flow rates $Q(t) \sim 0.1$ cm$^3$/s that are much larger than the steady flow rate $\sim 0.005$ cm$^3$/s corresponding to the continuous evacuation of the CSF produced in the ventricles \citep{linninger2016cerebrospinal}. The associated dynamics, involving complex nonlinear interactions between the fluid motion and the displacement of the soft tissues of the CNS, plays a fundamental role in the physiological function of CSF as a vehicle for metabolic-waste clearance as well as in the development of CNS diseases, such as idiopathic normal pressure hydrocephalus (iNPH) \citep{linninger2016cerebrospinal}. In-vivo measurements using non-invasive experimental methods based on magnetic resonance imaging (MRI) \citep{feinberg1987human} and advanced modelling studies employing computational-fluid-dynamics (CFD) techniques \citep{kurtcuoglu2007computational,gupta2009three,sweetman2011three} have been instrumental in increasing our understanding of the underlying fluid-structure interaction problem. Despite recent progress, many fundamental questions remain open, as summarized in a recent review by \cite{linninger2016cerebrospinal}.

Attention is focused here on the pulsating flow in the cerebral aqueduct, driven by the interventricular pressure difference $\Delta p(t) \simeq p_3-p_4$, where $p_3(t)$ and $p_4(t)$ are the time-dependent, nearly uniform \citep{kurtcuoglu2007computational} values of the pressure in the third and fourth ventricles. Approximate descriptions assuming fully developed unidirectional Womersley-like flow \citep{bardan2012simple} or a simplified hydraulic model \citep{longatti2019form} are available, as well as computational studies of the associated flow including realistic anatomical shapes  \citep{jacobson1996fluid,jacobson1999computer,fin2003three,kurtcuoglu2007computational}. In this manuscript we report on the development of a simplified model accounting for the relevant controlling parameters, to be used in predicting the relation between the interventricular pressure difference $\Delta p(t)=p_3-p_4$ and the resulting volume flow rate in the aqueduct $Q(t)$. Accurate knowledge of this relation is needed to quantify $\Delta p(t)$ from MRI measurements of CSF flow rate $Q(t)$ \citep{bardan2012simple}. Since the cerebral aqueduct is the narrowest interventricular passage, most of the pressure drop associated with the CSF motion in the ventricular system occurs as the CSF flows between the third and fourth ventricles \citep{sweetman2011three,bardan2012simple}. As a result, the value of $\Delta p(t) \simeq p_3-p_4$ provides an approximate representation for the so-called transmantle pressure \citep{jacobson1996fluid,jacobson1999computer}, the pressure difference between the lateral ventricles and the upper convexity of the SAS. Direct measurements of this quantity require very accurate simultaneous readings from two separate high-resolution pressure sensors \citep{vinje2019respiratory}, an invasive procedure with considerable risk factors \citep{penn2005pressure}, thereby fostering interest in indirect non-invasive techniques based on MRI measurements of $Q(t)$.

The interventricular pressure difference $\Delta p(t)=p_3-p_4$ and the resulting volume flow rate in the aqueduct  $Q(t)$ exhibit distinct quasi-periodic components associated with the cardiac and respiratory pulsations, with corresponding periods $T \sim 1\,$s and $T \sim 4-5\,$s, respectively. Except for one study~\citep{dreha2015inspiration}, all available MRI flow measurements  \citep{chen2015dynamics,takizawa2017characterization,yatsushiro2018visualization} indicate that the cardiac component of the flow velocity is somewhat larger than the respiratory component, that being also the case for the flow elsewhere in the cranial cavity \citep{yildiz2017quantifying} and along the spinal canal \citep{friese2004influence}, the latter flow displaying increasing effects of respiration on approaching the lumbar region. Since the pressure difference required to accelerate the flow in the aqueduct can be anticipated to be inversely proportional to the oscillation period, as follows from a balance between the local flow acceleration and the pressure force per unit mass, the interventricular pressure difference $\Delta p(t)$ associated with the cardiac cycle is much larger than that of the respiratory cycle, a conclusion supported by direct pressure measurements \citep{vinje2019respiratory}. By way of contrast, the stroke length $L_s$ (or stroke volume) of the oscillatory flow, linearly proportional to the oscillation period, is significantly larger for the respiratory-driven flow  \citep{vinje2019respiratory}, as shown by MRI measurements \citep{yatsushiro2018visualization,yamada2013influence}. As a result, studies focusing on the determination of the stroke volume, an important parameter characterizing aqueductal flow in iNPH patients and their response to shunting \citep{ringstad2015aqueductal,shanks2019aqueductal}, must account for the effects of respiration. The present analysis is general, in that the parametric ranges investigated include conditions corresponding to both cardiac and respiratory driven motion.

\section{Scales and order-of-magnitude estimates}

The cerebral aqueduct, shown in figure~\ref{fig:1}, is a slightly curved slender canal of length $L \sim 10-15$ mm and radius of curvature $R_c \sim 40-60$ mm. Its detailed anatomic shape is displayed  in the outer contour given in figure~\ref{fig:1}(c), measured in a healthy subject with MRI imaging techniques. The aqueduct's nearly cylindrical shape \citep{fin2003three} can be described by assuming
a circular section with slowly varying radius $a(s) \ll L$, with $s$ representing the distance along the centerline of the aqueduct measured from the third ventricle. The variation corresponding to the aqueduct of figure~\ref{fig:1}(c), obtained after smoothing the segmented contour, is shown in ~\ref{fig:1}(e). The corresponding aqueduct volume $\int_0^L \upi a^2 {\rm d} s$ can be equated to that of a circular cylinder with the same length $L$ to define the characteristic aqueduct radius $a_c$ from $\upi a_c^2 L=\int_0^L \upi a^2 {\rm d} s$, yielding typical values of order $a_c \sim 1-1.5 \, {\rm mm} \ll L$.

We address the pulsating motion induced by the periodic pressure difference $\Delta p(t) \simeq p_3-p_4$, resulting in a periodic volumetric flow rate $Q(t)$ with the same period $T$. The corresponding stroke volume $V_s=\frac{1}{2} \int_0^T |Q| {\rm d} t$ has been measured to be comparable to the aqueduct volume $\upi a_c^2 L$ \citep{ringstad2015aqueductal,markenroth2018investigation,shanks2019aqueductal}. Correspondingly, the characteristic stroke length $L_s=V_s/(\upi a_c^2)$ is comparable to the aqueduct length $L$. Since the temporal variations of the aqueduct volume, associated with the deformation of the bounding tissue, are much smaller than the aqueduct volume itself \citep{kurtcuoglu2007computational}, the aqueduct can be assumed to be rigid for the analysis of the flow, as done below in our analysis. In this respect, the problem is fundamentally different from that of CSF flow in the spinal canal \citep{linninger2016cerebrospinal}, where there exists close coupling between the fluid motion and the displacement of the canal walls, leading to a complex fluid-structure interaction problem that has been recently described with a linear elastic model adopted for the canal deformation \citep{sanchez2018bulk,Lawrence2019dispersion}. For the flow in the aqueduct, the errors associated with the use of the rigid-wall approximation can be anticipated to be on the order of the ratio of the cyclic variation of the aqueduct volume to the stroke volume, a quantity of the order of $10^{-2}$, as revealed by MRI brain-motion scans \citep{kurtcuoglu2007computational}.

The above estimates can be used to anticipate the character of the flow in the aqueduct, as done below. It is important to remark here that the analysis must consider the existence of three distinct regions, namely, the long central part of the aqueduct, where the flow is slender, and the two (much shorter) non-slender opening regions connecting the ends of the aqueduct with the ventricles.

Inside the aqueduct the flow is slender, with characteristic streamwise and transverse lengths $L$ and $a_c \ll L$. Since the streamlines are always nearly aligned inside the aqueduct, the transverse pressure variations are of order $(a_c/L)^2 \Delta p \ll \Delta p$ and thus can be neglected in the first approximation. The characteristic streamwise velocity is given by $U_c = \omega L_s \sim L_s/T$ in terms of the angular frequency $\omega =2 \upi/T$, yielding $U_c^2/L=\omega^2 L_s^2/L$ and $\omega U_c=\omega^2 L_s$ for the orders of magnitude of the convective and local accelerations, respectively, their relative importance being therefore measured by the parameter $L_s/L \sim 1$, the inverse of the relevant Strouhal number. The viscous time across the aqueduct $a_c^2/\nu \sim 1\,$s is comparable in magnitude to the flow oscillation period $T$, thereby yielding order-unity values of the Womersley number $\alpha=(\omega a_c^2/\nu)^{1/2}$; the associated Stokes number $\alpha^2$ representing the ratio of the magnitudes of the local acceleration to the viscous force (per unit mass). The order-of-magnitude analysis therefore reveals that  inside the aqueduct all terms in the streamwise momentum equation have comparable magnitude. Since convective acceleration has a non-negligible effect, the relation between $\Delta p$ and the flow rate $Q(t)$ is inherently nonlinear, thereby compromising the accuracy of studies adopting a presumed linear relation \citep{longatti2019form}. Also, analyses neglecting convective terms by assuming developed (i.e.\ Womersley-like) flow either everywhere \citep{bardan2012simple} or at the aqueduct entrance \citep{kurtcuoglu2007computational,vinje2019respiratory}, a valid approximation when $L_s/L\ll 1$, are necessarily inaccurate when $L_s \sim L$, the prevailing condition found in healthy and iNPH subjects \citep{ringstad2015aqueductal,markenroth2018investigation,shanks2019aqueductal}.

In the opening regions, of characteristic size $a_c$, the flow is non-slender, with characteristic velocity $U_c=\omega L_s$, corresponding to a Strouhal number $a_c/L_s \sim a_c/L \ll 1$ and a Reynolds number $U_c a_c/\nu \sim \alpha^2/(a_c/L) \gg 1$. Since local acceleration and viscous forces have small effects scaling with $a_c/L \ll 1$, the flow in the opening regions is quasi-steady and inviscid in the first approximation. As seen in previous CFD simulations of the flow in the third ventricle \citep{kurtcuoglu2007computational}, the resulting streamline pattern is very different for outflow, when the stream separates to form a jet that discharges into the ventricle, and for inflow, where the CSF accelerates from rest approaching the aqueduct entrance from all directions, with the sum of the pressure and the kinetic energy remaining constant along any given streamline. Since $L_s \sim L$, the associated pressure drop in the opening region, of order $\rho U_c^2 =\rho \omega^2 L_s^2$, is comparable in magnitude to the pressure drop along the aqueduct, of order $\rho U_c \omega L =\rho \omega^2 L_s L$, so that both contributions must be accounted for in evaluating the interventricular pressure $\Delta p(t)=p_3-p_4$ for a given volumetric flow rate $Q(t)$.

\section{Simplified description of the flow}

The aqueduct is seen as a slender canal connecting two large reservoirs whose pressure varies periodically in time. The problem will be posed as that of determining the interventricular pressure difference $\Delta p(t)$ that results in a given volume flow rate $Q(t)$, with the latter having a zero mean value, i.e.\ $\int_0^T Q {\rm d}t=0$. The slender-flow approximation $a_c/L \ll 1$ will be employed in simplifying the solution, with the dimensionless problem reducing to that of finding the pressure difference $\Pi=\Delta p/(\rho \omega U_c L)$ associated with a dimensionless flow rate $\bar{Q}=Q/(\omega V_s)$ for a given aqueduct anatomy, defined by the distribution of aqueduct radius $\bar{a}=a/a_c$, and given values of the controlling parameters $L_s/L \sim 1$ and $\alpha^2=\omega a_c^2/\nu \sim 1$.

The function $\bar{Q}$ must satisfy $\int_0^{2\upi} |\bar{Q}| {\rm d} \tau=2$, as follows from the definition of the stroke volume $V_s=\frac{1}{2} \int_0^T |Q| {\rm d} t$, with $\tau=\omega t$ representing a dimensionless time. For the cardiac-induced motion, the typical temporal variation $\bar{Q}(\tau)$ over a cycle is represented by the solid curve on the upper plot of figure~\ref{fig:2}, to be discussed later, corresponding to cardiac-gated MRI measurements of the aqueduct flow rate in a healthy subject. A Fourier analysis of the signal reveals that the first-mode, of period $2 \upi/\omega$, is dominant \citep{bardan2012simple}, so that for many quantitative purposes a simple sinusoidal function $\bar{Q}(\tau)=\tfrac{1}{2} \sin \tau$ can be used to represent the flow.

The computation of $\Pi(\tau)$ for a given $\bar{Q}(\tau)$ requires consideration of the flow both inside the aqueduct and in the opening regions connecting the aqueduct to the ventricles. The slender flow in the aqueduct is described in terms of the dimensionless streamwise distance from the third ventricle $x=s/L$ and the dimensionless radius $r$, the latter obtained by scaling the radial distance with the characteristic aqueduct radius $a_c$. Neglecting small terms of order $(a_c/L)^2$ and $a_c/R_c$ in writing the conservation equations inside the aqueduct leads to the axisymmetric boundary-layer problem
\begin{align}
\frac{\p u}{\p x}+\frac{1}{r} \frac{\p}{\p r} \left(r v \right)&=0, \label{cont_eq} \\
\frac{\p u}{\p \tau}+\frac{L_s}{L} \left(u \frac{\p u}{\p x}+v \frac{\p u}{\p r}\right)&=-\frac{\p p'}{\p x} +\frac{1}{\alpha^2} \frac{1}{r} \frac{\p}{\p r} \left(r \frac{\p u}{\p r}\right), \label{mom_eq}
\end{align}
where the dimensionless streamwise and radial velocity components $u$ and $v$ are scaled with $U_c=\omega L_s$ and $U_c a_c/L$, respectively. The axial velocity must satisfy $\bar{Q}(\tau)=\int_0^{\bar{a}} 2 r u {\rm d} r$, as follows from the selected scaling. The streamwise pressure gradient $P_x(x,\tau)=\p p'/\p x$, where $p'$ denotes the spatial pressure difference scaled with $\rho \omega U_c L$, is unknown and must be determined as part of the integration.

Equations~\eqref{cont_eq} and~\eqref{mom_eq} must be integrated for $0 \le x \le 1$ and $0 \le r \le \bar{a}(x)$ subject to the boundary conditions
\begin{equation} \label{bc1}
\frac{\p u}{\p r}=v=0 \quad {\rm at} \quad r=0 \qquad {\rm and} \qquad u=v=0 \quad {\rm at} \quad r=\bar{a}(x).
\end{equation}
To write the needed boundary conditions for $u$ at the two ends of the canal $x=0,1$ consideration must be given to the CSF motion in the near-field region, corresponding to distances from the canal opening of order $a_c \ll L$, where the flow is non-slender, with characteristic velocities of order $U_c$. Using $a_c$ and $U_c$ as characteristic scales of length and velocity, reduces the momentum equation to
\begin{equation}
\left(\frac{a_c}{L}\right) \frac{\p \mathbf{v}}{\p \tau}+\frac{L_s}{L}  \mathbf{v} \cdot \nabla \mathbf{v}=-\nabla p'+\frac{1}{\alpha^2} \left(\frac{a_c}{L}\right) \nabla^2 \mathbf{v}. \label{mom_entrance}
\end{equation}
This dimensionless equation reveals that, in the limit $a_c/L \ll 1$ considered here, with $\alpha \sim 1$ and $L_s/L \sim 1$, the flow in the opening regions is quasi-steady and nearly inviscid in the first approximation. The resulting streamline pattern, shown in Fig.~5 of \cite{kurtcuoglu2007computational}, is drastically different for inflow (i.e.\ $\bar{Q}>0$ at $x=0$ or $\bar{Q}<0$ at $x=1$) and outflow (i.e.\ $\bar{Q}<0$ at $x=0$ or $\bar{Q}>0$ at $x=1$). For outflow, the stream separates to form a jet that discharges into the ventricle, with the pressure across the jet being approximately equal to that of the ventricle. In the boundary-layer approximation employed here in describing the flow inside the aqueduct, no boundary condition is needed for the flow velocity at the canal end when outflow is present.

For inflow, on the other hand, the CSF accelerates from rest approaching the aqueduct entrance from all directions. As follows from the steady inviscid form of~\eqref{mom_entrance}, the stagnation pressure in the opening region, outside from a thin near-wall viscous boundary layer, is equal to the pressure in the feeding reservoir. Since the streamlines align on entering the aqueduct, the pressure is uniform across the entire entrance section, so that the condition $p'+(L_s/L) |\mathbf{v}|^2/2=$ constant implies that the velocity must also be uniform there, thereby leading to the alternating boundary conditions
\begin{equation}\label{bc2}
\left\{ \begin{array}{lll} \bar{Q}>0: & u=\bar{Q}(\tau)/\bar{a}^2(0) & {\rm at} \; x=0 \\ \bar{Q}<0: & u=\bar{Q}(\tau)/\bar{a}^2(1) & {\rm at} \; x=1 \end{array} \right.,
\end{equation}
involving the dimensionless local radii $\bar{a}(0)=a(0)/a_c$ and $\bar{a}(1)=a(L)/a_c$ at the two aqueduct's ends. Correspondingly, the pressure drop between the ventricle and the entrance of the aqueduct is $(L_s/L)[\bar{Q}/\bar{a}(0)]^2/2$ if $\bar{Q}>0$ and $(L_s/L)[\bar{Q}/\bar{a}(1)]^2/2$ if $\bar{Q}<0$, as follows from conservation of stagnation pressure.

As revealed by~\eqref{mom_entrance}, the assumption of quasi-steady flow in the entrance region, valid over most of the cycle, can be expected to fail when $\bar{Q}$ vanishes, during short flow-reversal stages of relative duration $\Delta \tau \simeq a_c/L_s$ when the velocity is of order $a_c/L_s$. As a result, the local acceleration becomes comparable to the convective acceleration in the entrance region, while viscous forces are still negligible there. During this short stage the flow is inviscid also inside the aqueduct, where the momentum balance~\eqref{mom_eq} reduces to ${\p u}/{\p \tau}=-\p p'/\p x$, which can be integrated across the section to show that $-\p p'/\p x={\rm d} \bar{Q}/{\rm d}\tau |_0$, involving the rate of variation of the flow rate at the instant of flow reversal ${\rm d} \bar{Q}/{\rm d}\tau |_0$. Integrating this last equation shows that, during this short stage, the pressure drop along the aqueduct is given by ${\rm d} \bar{Q}/{\rm d}\tau |_0 \sim 1$, while the corresponding pressure drop across the entrance region is small, of order $a_c/L_s$.

For given values of $L_s/L$ and $\alpha$, a given aqueduct shape $\bar{a}(x)$, and a given 2$\upi$-periodic dimensionless flow rate $\bar{Q}(\tau)$, integration of~\eqref{cont_eq} and~\eqref{mom_eq} subject to the boundary conditions stated in~\eqref{bc1} and~\eqref{bc2} determines the velocity field $u(x,r,\tau)$ and $v(x,r,\tau)$ and associated pressure gradient $P_x(x,\tau)$. As previously explained, the interventricular pressure difference $p_3-p_4$ is the sum of the pressure loss along the slender portion of the aqueduct and the pressure loss at the aqueduct entrance, the latter evaluated earlier, below~\eqref{bc2}, with use of Bernoulli's law. In our dimensionless formulation, the result can be expressed in the form
\begin{equation}
\Pi(\tau)=\frac{p_3-p_4}{\rho \omega^2 L_s L}=-\int_0^1 P_x {\rm d}x \left\{ \begin{array}{lll} +\tfrac{1}{2} \frac{L_s}{L} \frac{\bar{Q}^2}{\bar{a}^4(0)} & {\rm if} & \bar{Q}> 0 \\ -\tfrac{1}{2} \frac{L_s}{L} \frac{\bar{Q}^2}{\bar{a}^4(1)} & {\rm if} & \bar{Q}< 0 \end{array} \right.. \label{Pi(tau)}
\end{equation}
This result is to be compared with the pressure drop
\begin{equation}
\Pi(\tau)=\frac{8 \bar{Q}}{\alpha^2} \int_0^1 \frac{{\rm d}x}{\bar{a}^4(x)}
\label{eq:Pois}
\end{equation}
corresponding to the quasi-steady Poiseuille velocity profile $u=-\tfrac{1}{4} \alpha^2 P_x (\bar{a}^2-r^2)$, obtained in the present formulation when taking the limit $\alpha \ll 1$. Additional closed-form analytical solutions can be found in the inviscid limit $\alpha \gg 1$, when~\eqref{Pi(tau)} can be seen to reduce to
 \begin{equation}
\Pi(\tau)=\left(\int_0^1 \frac{{\rm d}x}{\bar{a}^2(x)}\right)  \frac{{\rm d} \bar{Q}}{{\rm d} \tau} \, \left\{ \begin{array}{lll} +\tfrac{1}{2} \frac{L_s}{L} \frac{\bar{Q}^2}{\bar{a}^4(1)} & {\rm if} & \bar{Q}> 0 \\ -\tfrac{1}{2} \frac{L_s}{L} \frac{\bar{Q}^2}{\bar{a}^4(0)} & {\rm if} & \bar{Q}< 0 \end{array} \right., \label{Pi(tau)2}
\end{equation}
 and also for $L_s/L \ll 1$, when convective terms have a small effect on the aqueduct flow, as can be inferred from~\eqref{mom_eq}, resulting in a linear Womersley-like problem that can be solved explicitly using a complex Fourier series representation for the flow rate
 \begin{equation}
 \bar{Q}(\tau)={\rm Re}\left(\sum_{n=1}^{\infty} Q_n \ue^{\ui n \tau} \right)
 \end{equation}
 to give $P_x={\rm Re}\left(\sum_{n=1}^{\infty} A_n \ue^{\ui n \tau} \right)$,
 where
 \begin{equation}
 A_n(x)=-\frac{\ui \, n Q_n}{\bar{a}^2(x)}\left(1+\frac{J_1(\beta_n)}{(\beta_n/2)J_0(\beta_n)-J_1(\beta_n)}\right) \quad {\rm and} \quad \beta_n(x)=\frac{\ui-1}{\sqrt{2}} \sqrt{n} \alpha \bar{a}(x)
 \end{equation}
 are complex functions that vary along the aqueduct, with Re() denoting the real part of a complex expression and $J_0$ and $J_1$ representing the Bessel functions of order $0$ and $1$, respectively. Since the pressure drop at the aqueduct entrance becomes negligibly small for $L_s/L \ll 1$, the dimensionless interventricular pressure difference~\eqref{Pi(tau)} reduces in this case to
 \begin{equation}
 \Pi(\tau)={\rm Re}\left[\ui \sum_{n=1}^{\infty} n Q_n \ue^{\ui n \tau} \int_0^1 \frac{1}{\bar{a}^2} \left(1+\frac{J_1(\beta_n)}{(\beta_n/2)J_0(\beta_n)-J_1(\beta_n)}\right) {\rm d} x  \right]. \label{wom}
 \end{equation}

\section{Selected numerical results}

The governing equations were discretized using a Krause zig-zag finite-difference scheme in $x$ and $t$ \citep{Tannehill.etal.1997} (uniform $\Delta x = 1/200$; adaptive time step with mean $\Delta \tau = 2\upi/800$), combined with Chebyshev spectral collocation in $r$ (32 points). At each step in time, the equations are marched from $x=0$ to $1$ when $Q>0$, and from $x=1$ to $0$ when $Q<0$. The nonlinearity in the convective term is handled by an iterative fixed-point procedure at every marching step in $x$.
The computation was run in time until a 2$\upi$-periodic solution was reached, with convergence occurring after about 5--10 cycles. Special attention was given to the transition between outflow and inflow, occurring once at each end of the aqueduct during the flow cycle. 
In the proposed scheme, the velocity profile at the canal end where outflow is present is computed as part of the boundary-layer computation, with the corresponding velocity at the other end given by~\eqref{bc2}. As previously discussed, this approximation can be expected to fail during the short stages of flow reversal, as the quasi-steady approximation breaks down in the opening region. This was apparent in the numerical integrations, which revealed that, when $\bar{Q}$ vanishes, the resulting outflow velocity, although very small (typical peak values not exceeding $10^{-1}$), was not exactly zero, leading to a discontinuity in the temporal evolution at the canal end when switching between the boundary conditions in~\eqref{bc2}. This was accounted for in the numerical integration by incorporating a short transition stage, with duration $\Delta \tau \ll 1$ following the change of sign of $\bar{Q}$, during which the inflow velocity profile was continuously adapted with a linear temporal fit from that found at the end of the outflow period to the inflow uniform value defined in~\eqref{bc2}. The resulting value of $\Pi(\tau)$ was found to be independent of $\Delta \tau$ provided that $10^{-2} \ll \Delta \tau \ll 10^{-1}$.

\begin{figure}
\centering
\includegraphics[width=0.78\textwidth]{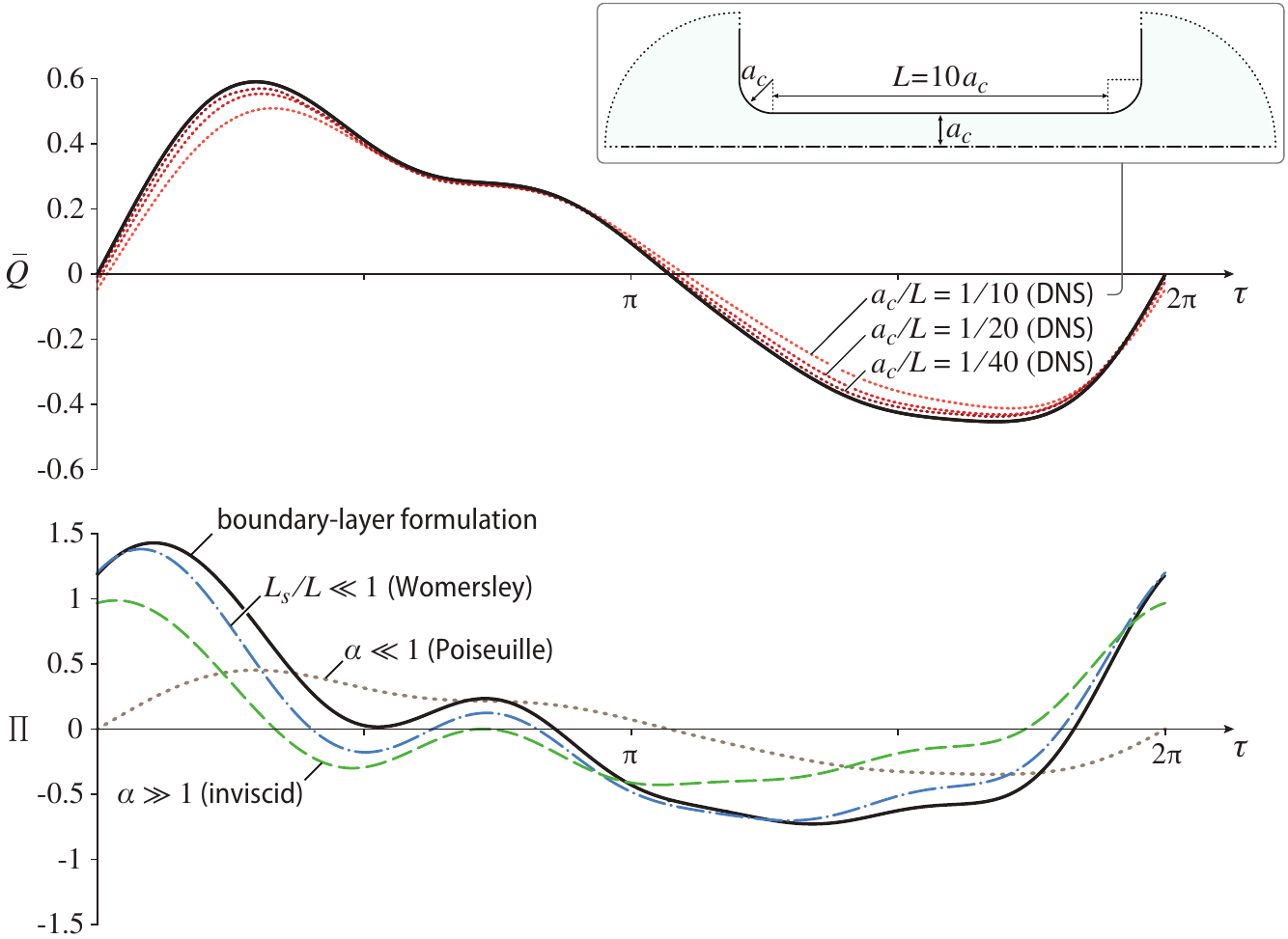}
\caption{The solid curve in the lower plot represents the dimensionless pressure difference $\Pi(\tau)$ determined from the simplified model with $L_s/L=0.93$, $\alpha=3.23$, and $\bar{a}=1$ for the  dimensionless function $\bar{Q}(\tau)$ represented by the solid curve in the upper plot, corresponding to the Fourier decomposition of MRI measurements of the volumetric flow rate in the aqueduct of a healthy 36-year old male subject using phase contrast \citep{Kevin}.
The additional dot-dashed, dashed, and dotted curves in the lower plot represent, respectively, the results of the Womersley approximation~\eqref{wom}, of the inviscid approximation~\eqref{Pi(tau)2}, and of the quasi-steady Poiseuille flow~\eqref{eq:Pois}. The dotted curves in the upper plot are obtained from direct numerical simulations (DNS) using the interventricular pressure difference $\Pi(\tau)$ represented by the solid curve in the lower plot for different values of $a_c/L$.}
\label{fig:2}
\end{figure}

The bottom plot in figure~\ref{fig:2} shows illustrative results corresponding to a canal of constant radius $\bar{a}=1$. The shape $\bar{Q}$ of the specific flow rate employed in this computation, shown as a solid curve in the upper plot, as well as the values of $L_s/L=0.93$, $\alpha=3.23$ correspond to those obtained using cardiac-gated MRI measurements of the aqueduct flow in a healthy human subject \citep{Kevin}. The periodic function $\Pi(\tau)$ evaluated from~\eqref{Pi(tau)} is shown as a solid curve. Because of the effect of the nonlinear convective terms, the average interventricular pressure, identically zero in the linear limit $L_s/L \ll 1$, takes a small non-zero value $\int_\tau^{\tau+2\upi} \Pi {\rm d} \tau/(2\upi)=0.04$, in agreement with previous findings~\citep{stephensen2002there}. For completeness, the figure also includes the pressure predictions obtained with Poiseuille flow and also in the two limits $L_s/L \ll 1$ (Womersley) and $\alpha \gg 1$ (inviscid). As can be seen, for this specific case the former limit, neglecting nonlinear terms while retaining the local acceleration, provides a largely satisfactory description of the interventricular pressure, with quantitative departures remaining below 20\% over most of the cycle. In contrast, the quasi-steady Poiseuille solution leads to severe underpredictions of interventricular pressure difference.

The dimensionless interventricular pressure shown in figure~\ref{fig:2} can be expressed in dimensional form with use of $\Delta p=\rho \omega^2 L_s L \Pi$. Using in the evaluation $L=10$ mm for the aqueduct length along with the standard cardiac frequency $\omega=2 \upi$ s$^{-1}$ reveals that the dimensionless peak value $\Pi \simeq 1.5$ in figure~\ref{fig:2} corresponds to an overpressure $\Delta p=p_3-p_4 \simeq 5.5$ Pa, consistent with existing measurements \citep{eide2010ventriculomegaly,vinje2019respiratory} and computations \citep{sweetman2011three} of instantaneous spatial pressure variations in the cranial cavity. As expected, the corresponding maximum overpressure predicted with Poiseuille velocity $p_3-p_4 \simeq 1.66$ Pa, corresponding to the peak $\Pi=0.452$ in figure~\ref{fig:2}, is significantly smaller.

Direct numerical simulations were used to test the accuracy of the simplified model. The computations considered the geometry illustrated in the inset of figure~\ref{fig:2}, corresponding to a duct of length $L$ and constant radius $a_c$ connecting two quasi-infinite reservoirs, with the smooth convex surface connecting the pipe with the reservoir having radius $a_c$. The axisymmetric Navier-Stokes equations were integrated for different values of $a_c/L$ using as boundary condition the interventricular pressure difference $\Pi(\tau)$ shown as a solid curve in the lower plot of figure~\ref{fig:2}. Resulting flow rates $\bar{Q}(\tau)$ are represented in the upper plot. As can be seen, the results rapidly converge to the original flow rate used in the simplified model, with relative differences scaling approximately with $a_c/L$.

\begin{figure}
\centering
\includegraphics[width=0.85\textwidth]{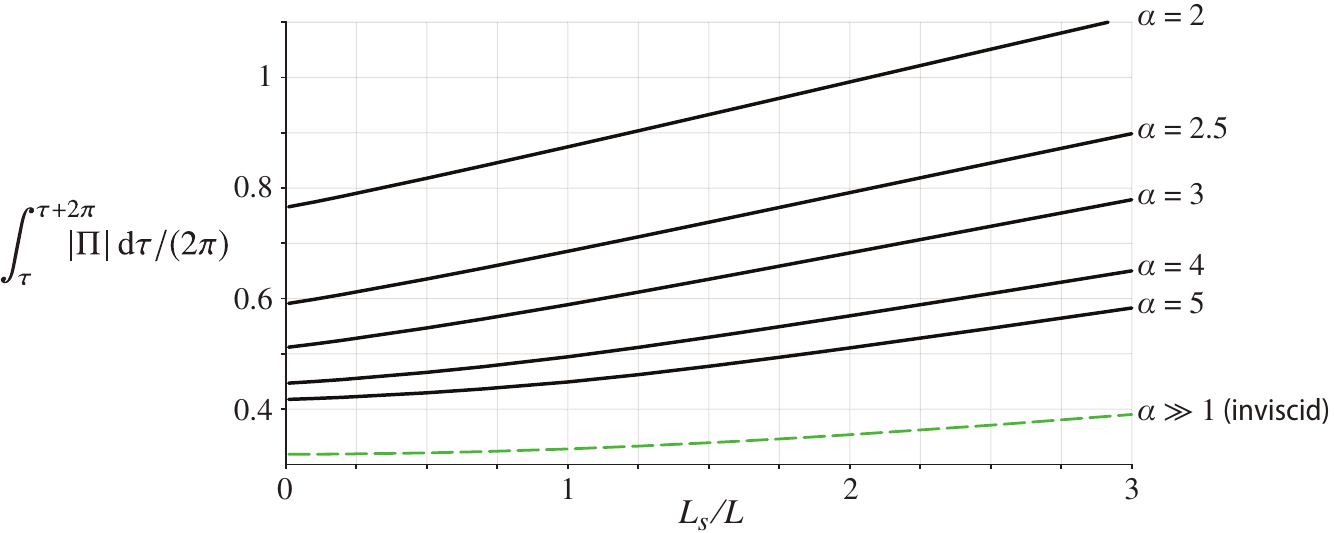}
\caption{The variation of $\int_\tau^{\tau+2\upi} |\Pi| {\rm d} \tau/(2\upi)$ with $L_s/L$ obtained from the simplified flow model for  $\bar{a}=1$, $\bar{Q}(\tau)=\tfrac{1}{2} \sin \tau$, and different values of $\alpha$.}
\label{fig:3}
\end{figure}

The model was used to quantify the parametric dependence of $\int_\tau^{\tau+2\upi} |\Pi| {\rm d} \tau/(2\upi)$, a measure of the oscillating force exerted on the brain. Results corresponding to $\bar{a}=1$ and $\bar{Q}(\tau)=\tfrac{1}{2} \sin \tau$ are plotted in figure~\ref{fig:3} as a function of $L_s/L$ for different values of $\alpha$, including the inviscid result $\int_\tau^{\tau+2\upi} |\Pi| {\rm d} \tau/(2\upi)=1/\upi+(L_s/L)/16$ corresponding to the limit $\alpha \gg 1$. The figure covers the range of conditions typically found in healthy subjects, characterized by values of $L_s/L$ of order unity and Womersley numbers ranging from $\alpha \simeq 2$ for the respiratory cycle to $\alpha \simeq 4$ for the cardiac cycle. The plot can also be used in connection with iNPH patients, who typically show enlarged aqueducts and higher tidal volumes \citep{ringstad2015aqueductal,markenroth2018investigation,shanks2019aqueductal}, corresponding to larger values of the parameters $\alpha$ and $L_s/L$.

\section{Concluding remarks}

The simplified flow model presented above can be instrumental in developing protocols for non-invasive patient-specific quantification of the transmantle pressure difference between the lateral ventricles and the SAS from MRI measurements of the aqueduct radius $a(s)$ and volumetric flow rate $Q(t)$. These models have the potential for improving our current understanding of intracranial flow dynamics, associated with the development of CNS diseases, enabling the development of early-diagnosis techniques.

Declaration of Interests. The authors report no conflict of interest.

\begin{acknowledgments}
We thank Dr. Kevin King and Dr. Victor Haughton for insightful discussions, and the former and Ms. Annie Malekie, Ms. Ke Wei, and Ms. Thao Tran at the Huntington Medical Research Institutes for providing the MRI measurements used in figures~\ref{fig:1} and~\ref{fig:2}. The work of ALS was supported by the National Science Foundation through grant \# 1853954. The work of WC was supported by the `Convenio Plurianual Comunidad de Madrid -- Universidad Carlos III de Madrid' through grant CSFLOW-CM-UC3M.
\end{acknowledgments}

\bibliographystyle{jfm}
\bibliography{references}

\end{document}